\documentclass[a4paper,11pt]{desyprocA42}
\usepackage[utf8]{inputenc}

\usepackage{graphicx}
\usepackage{amssymb}
\usepackage{epstopdf}
\usepackage{amsmath}
\usepackage{array}
\usepackage{slashed}
\usepackage{hyperref}
\usepackage{color}
\usepackage{braket}
\usepackage[english]{babel}
\usepackage[parfill]{parskip}
\usepackage{float}
\usepackage[font={it}]{caption}

\newcommand{\beqn}{\begin{equation}}
\newcommand{\eeqn}{\end{equation}}
\newcommand{\degree}{\ensuremath{^\circ}}

\newcommand{\LSP}{\widetilde{\chi}_1^0}
\newcommand{\neutralinotwo}{\widetilde{\chi}_2^0}
\newcommand{\neutralinothree}{\widetilde{\chi}_3^0}
\newcommand{\neutralinofour}{\widetilde{\chi}_4^0}

\newcommand{\charginoone}{\widetilde{\chi}_1^{\pm}}
\newcommand{\charginotwo}{\widetilde{\chi}_2^{\pm}}
\newcommand{\stau}{\widetilde{\tau}}
\newcommand{\stauone}{\widetilde{\tau}_1}
\newcommand{\stautwo}{\widetilde{\tau}_2}

\newcommand{\seL}{\widetilde{e}_L}
\newcommand{\seR}{\widetilde{e}_R}
\newcommand{\smuL}{\widetilde{\mu}_L}
\newcommand{\smuR}{\widetilde{\mu}_R}
\newcommand{\snuel}{\widetilde{\nu}_e}
\newcommand{\snumu}{\widetilde{\nu}_{\mu}}
\newcommand{\snutau}{\widetilde{\nu}_{\tau}}

\newcommand{\chargino}{\widetilde{\chi}^{\pm}}

\newcommand{\GeV}{\text{ GeV}}

\newcommand{\pkg}[1]{\texttt{#1}}

\usepackage{url}
\usepackage[affil-it]{authblk}

\title{Dark matter relic density from observations of \newline supersymmetry at the ILC}
\author{Suvi-Leena Lehtinen$^{1,2}$\footnote{Talk presented at the International Workshop on Future Linear Colliders (LCWS15), Whistler, Canada, 2-6 November 2015. \\ email: suvi-leena.lehtinen@desy.de} , Mikael Berggren$^1$, Jenny List$^1$}
\affil{$^1$DESY, Notkestr. 85, 22607 Hamburg, Germany \\
$^2$University of Hamburg, Physics Department, Luruper Chaussee 149, 22761 Hamburg, Germany}
\date{26.2.2016}

\begin{document}

\maketitle

\begin{abstract}
Supersymmetry can explain the observed dark matter relic density with a neutralino dark matter particle and a coannihilating, almost mass-degenerate sparticle. If this were the case in nature, a linear electron positron collider like the ILC could discover the two sparticles if their masses are in the kinematic reach of the collider. This contribution discusses which observations are necessary at the ILC for predicting the dark matter relic density correctly and for confirming that the observed lightest neutralino is the only kind of dark matter. We take the case of stau coannihilation as an example.
\end{abstract}

\section{Introduction}

The Planck mission has made the most precise measurement of the dark matter relic density $\Omega_{CDM}$ so far. The Planck 2015 result is $\Omega_{CDM}=0.1197 \pm 0.0022$, which was measured from the temperature power spectrum of the cosmic microwave background and polarisation data \cite{Ade:2015xua}. The relic density could be explained by supersymmetric dark matter, where the lightest neutralino $\LSP$ is the dark matter candidate. In most sections of SUSY parameter space another almost mass-degenerate particle is required to bring the relic density to the observed range. This type of scenario is called a coannihilation scenario, and the coannihilating particle could be a $\stau$, $\widetilde{t}$ or a $\chargino$.

Coannihilation scenarios can be discovered at an electron-positron collider like the International Linear Collider (ILC) because nearly all events can be recorded and all particles can be reconstructed. If all the SUSY particles are discovered and their mixings measured, then a predicted value of the relic density can be calculated. If only part of the spectrum can be measured, can the relic density still be calculated correctly? Which particles need to be measured and what precision is needed to rule out contributions from other types of dark matter to the relic density? To answer these questions a $\stau$ coannihilation scenario was studied. This contribution discusses which measurements need to be made at the ILC to predict the relic density in the observed range. Section 2 introduces the stau coannihilation scenario. Section 3 presents the possibility of deducing the relic density from the ILC observations. Section 4 summarises the study.

\section{Choice of benchmark}

\begin{figure}[ht!]
\begin{minipage}[b]{0.45\linewidth}
\centering
\includegraphics[width=\textwidth]{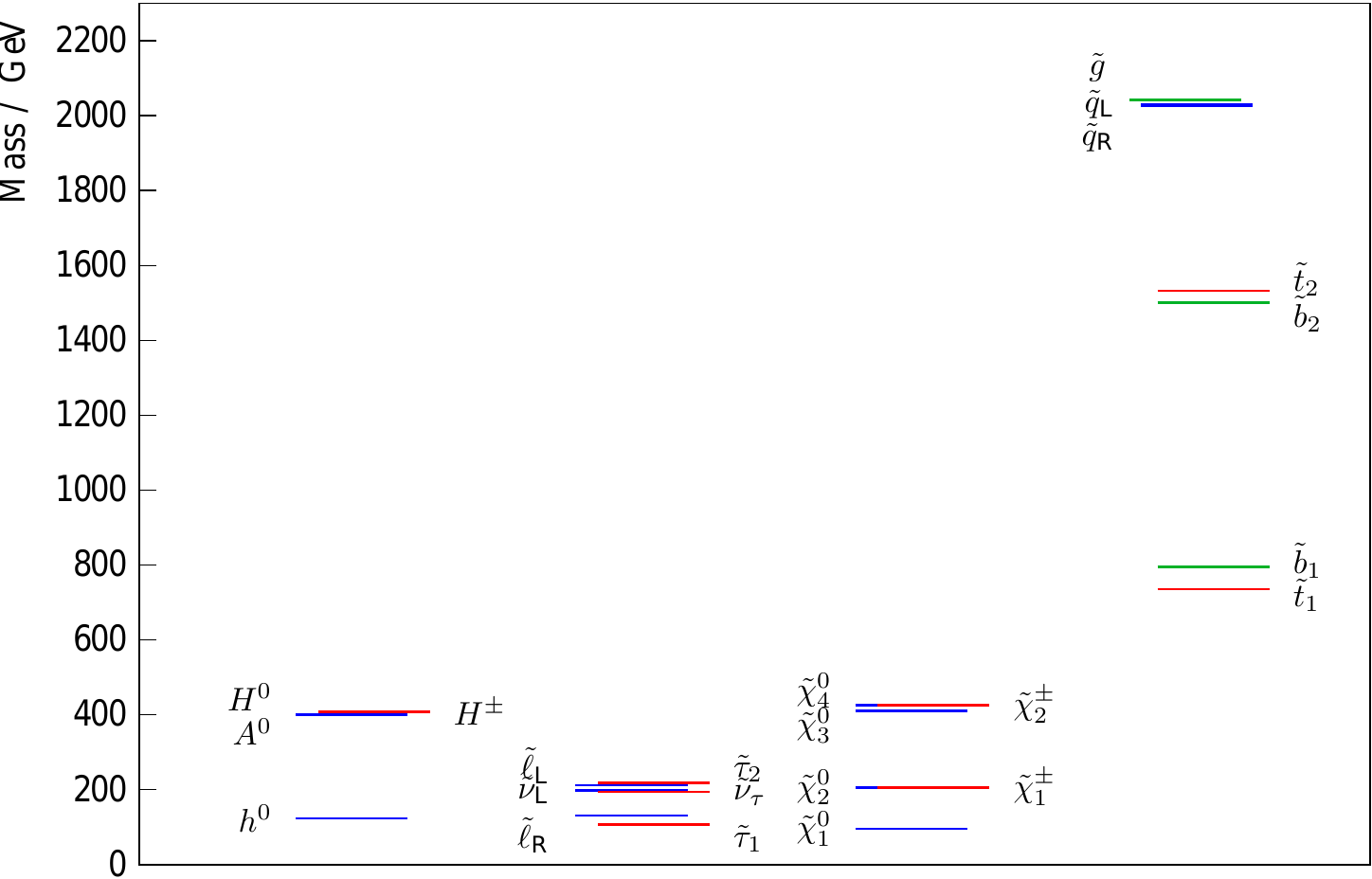}
\caption{STC8 spectrum}
\label{fig:STC8spectrumbig}
\end{minipage}
\hspace{0.5cm}
\begin{minipage}[b]{0.45\linewidth}
\centering
\includegraphics[width=\textwidth]{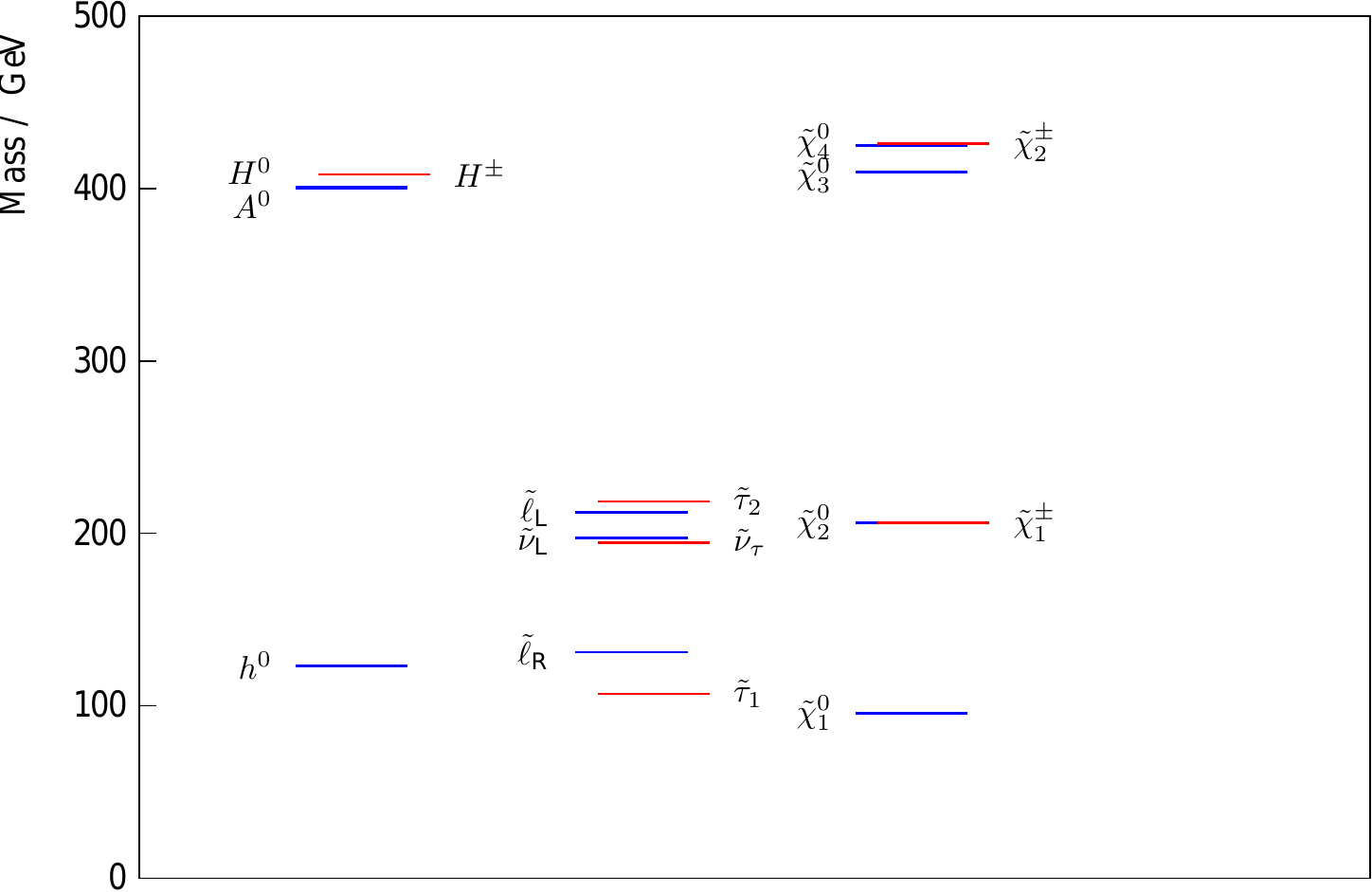}
\caption{Zoom of STC8 spectrum}
\label{fig:STC8spectrumsmall}
\end{minipage}
\end{figure}

Coannihilating particles can easily escape detection at the LHC due to their soft decay products. The ILC, however, is perfectly suited for discovering particles with small mass differences. A case well studied is stau coannihilation, specifically a scenario called STC8 \cite{Baer:2013ula}, which is a 12 parameter point in the pMSSM. Its parameters are listed in Table \ref{table:STC8parameters} and its spectrum is shown in Figs. \ref{fig:STC8spectrumbig} and \ref{fig:STC8spectrumsmall}. All sleptons, sneutrinos, $\LSP$, $\neutralinotwo$ and $\charginoone$ would be pair produced at the ILC with $\sqrt{s}= 500 $ GeV. Upgrading to $\sqrt{s}=1$ TeV would give access to the heavy Higgses, $\neutralinothree$, $\neutralinofour$ and $\charginotwo$.

The masses of the LSP and the light stau are 96 GeV and 107 GeV respectively, yielding a mass difference of 11 GeV. The composition of the LSP is $\LSP= -0.99\widetilde{B} + 0.03\widetilde{W}^0 -0.16 \widetilde{H}^0_d + 0.04\widetilde{H}^0_u$ and the stau mixing angle $\theta_{\tau}=71\degree$. The relic density has a value $\Omega_{CDM}= 0.113$, which we will take as the true value. The channels which contribute to the relic density the most are $\LSP \LSP \to l^+l^-$ with 73\%, $\LSP \stauone \to \gamma \tau$ with 16\% and $\LSP \stauone \to Z \tau$ with 3\% contributions.

Because many particles have masses near the LSP mass, there are many potential contributors to the relic density. In addition, the rich particle spectrum at the ILC challenges the analysis methods. An analysis of this scenario has been conducted in \cite{Berggren:2015qua} using the fast detector simulation of ILD and polarised electron and positron beams. The analysis finds that both the LSP mass and $\stauone$ mass can be measured with a 100 MeV or one permille accuracy with $\mathcal{L}= 500$ fb$^{-1}$ and polarisation $\mathcal{P}(e^-,e^+)=(80\%,-30\%)$. The masses of all other kinematically accessible particles can be measured with precisions of 1-5\%.

\begin{table}[ht!]{}
\centering
\begin{tabular}{ | c | c | }
	\hline						
 parameter &  model value 	 	 \\ \hline
	$\tan \beta$  & 10      \\ 
 	$\mu$ &	400	GeV	  	 \\ 
	$m_A$ & 400	GeV	    \\ 
 	$A_{\tau, t, b}$ &	-2400	GeV	  	 \\ 
 	$M_E$ &	119	GeV	 \\ 
  	$M_L$ &	205	GeV	   \\ 
 	$M_Q(1,2)$ &	2000 	GeV	  	 \\ 
 	$M_Q(3)$ &	1500	GeV	  	 \\ 
 	$M_U(1,2)=M_D(1,2)$ &	2000	GeV	  	   \\ 
 	$M_U(3)$=$M_D(3)$  &	 800	GeV	  	   \\ 
  	$M_1$ &	 100	GeV	  	   \\ 
 	$M_2$ &	 210	GeV	  	   \\ 
 	$M_3$ &	 2000	GeV	  	   \\ 

 	$m_{t}$ & 173.1	GeV	   \\ \hline
\end{tabular}
\
\caption{STC8 SUSY parameters}
\label{table:STC8parameters}
\end{table}

\section{Relic density from ILC measurements}

It is possible to calculate the relic density for a full SUSY model with \pkg{micrOMEGAs2.4.5} \cite{Belanger:2006is}. This can be done with SUSY parameters or with the SUSY observables via a file in the Les Houches format. We follow the latter approach and hence a full SUSY spectrum is needed as an input to \pkg{micrOMEGAs}. In the first step, only some masses and mixings were varied and the rest were fixed to their true values. Later the unobserved particle properties were treated as nuisance parameters and varied uniformly. 

\begin{figure}[ht!]
\begin{center}
\leavevmode
\includegraphics[width=0.6\textwidth]{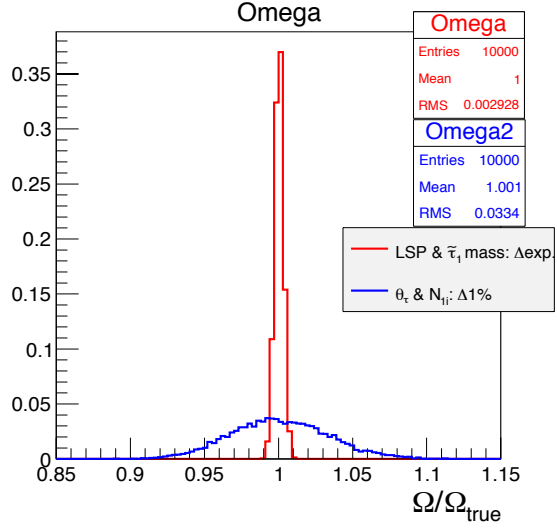}
\caption{Distribution of predicted $\Omega$ when LSP mass and $\stauone$ endpoint are varied (red) and when $\LSP$ and $\stauone$ mixings are varied (blue). The mixings need to be measured to 1\% to get a 3.3\% uncertainty on $\Omega$.}
\label{fig:LSPstau-massesvsmixings}
\end{center}
\end{figure}

Let us first consider the LSP and $\stauone$ properties only. The mass of the LSP and the value of the $\stauone$ endpoint can be measured to the permille level at the ILC with $\sqrt{s}=500$ GeV, $\mathcal{L}=500$ fb$^{-1}$ and polarisation $\mathcal{P}$(+80,-30): 0.15\% for the LSP mass and 0.24\% for the $\stauone$ endpoint or, equivalently, a 0.16\% uncertainty on the mass of the $\stauone$. If the LSP mass and the $\stauone$ endpoint are varied according to Gaussian distributions with the respective experimental uncertainties, then the calculated relic density has a narrow distribution centred around the true delic density, as is shown in Fig. \ref{fig:LSPstau-massesvsmixings}. The root mean square of the distribution is 0.3\%. The same figure shows the relic density distribution if the stau mixing angle $\theta_{\tau}$ and neutralino mixing matrix elements $N_{11}, N_{12}, N_{13}, N_{14}$, are all varied by 1\%. This results in a 3.3\% uncertainty on the relic density, which is of the same order as the Planck precision. However, an uncertainty of one percent for the mixings is ambitious: $\theta_{\tau}$ could be determined from the polarised production cross sections of $\stauone$ pairs \cite{Boos:2003vf} which could be measured to a three percent uncertainty with $\mathcal{L}=500$ fb$^{-1}$ in a similar scenario \cite{Bechtle:2009em}. If the mixings were measured to 2\% accuracy, then the uncertainty on the relic density would double.

\begin{figure}[ht!]
\begin{center}
\leavevmode
\includegraphics[width=0.6\textwidth]{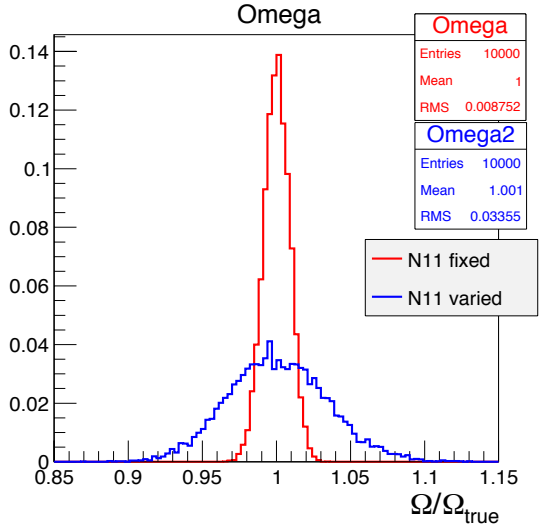}
\caption{The LSP and $\stauone$ masses and mixings varied together (blue curve). If $N_{11}$ is fixed (red curve), the RMS shrinks to about one third. In other words, the uncertainty on the main coupling of the LSP accounts for two thirds of the uncertainty on $\Omega$.}
\label{fig:LSPN11driving}
\end{center}
\end{figure}

If the LSP and $\stauone$ masses and mixings are varied simultaneously, the uncertainty on the relic density is 3.4\% (blue curve in Fig. \ref{fig:LSPN11driving}) compared with 3.3\% if only the mixings are varied (blue curve in Fig. \ref{fig:LSPstau-massesvsmixings}). Hence the mixings dominate the uncertainty. It is particularly important to measure the largest mixing matrix element $N_{11}$ of the LSP as precisely as possible. This can be seen from Fig. \ref{fig:LSPN11driving}, where the effect of varying the value of $N_{11}$ is shown. If $N_{11}$ is fixed to its model value, then the uncertainty on the relic density is reduced to approximately a third. It has not been investigated yet how precisely the mixing matrix elements can be measured at the ILC, although a method was presented in \cite{Nojiri:1994it}.

\begin{figure}[ht!]
\begin{center}
\leavevmode
\includegraphics[width=0.6\textwidth]{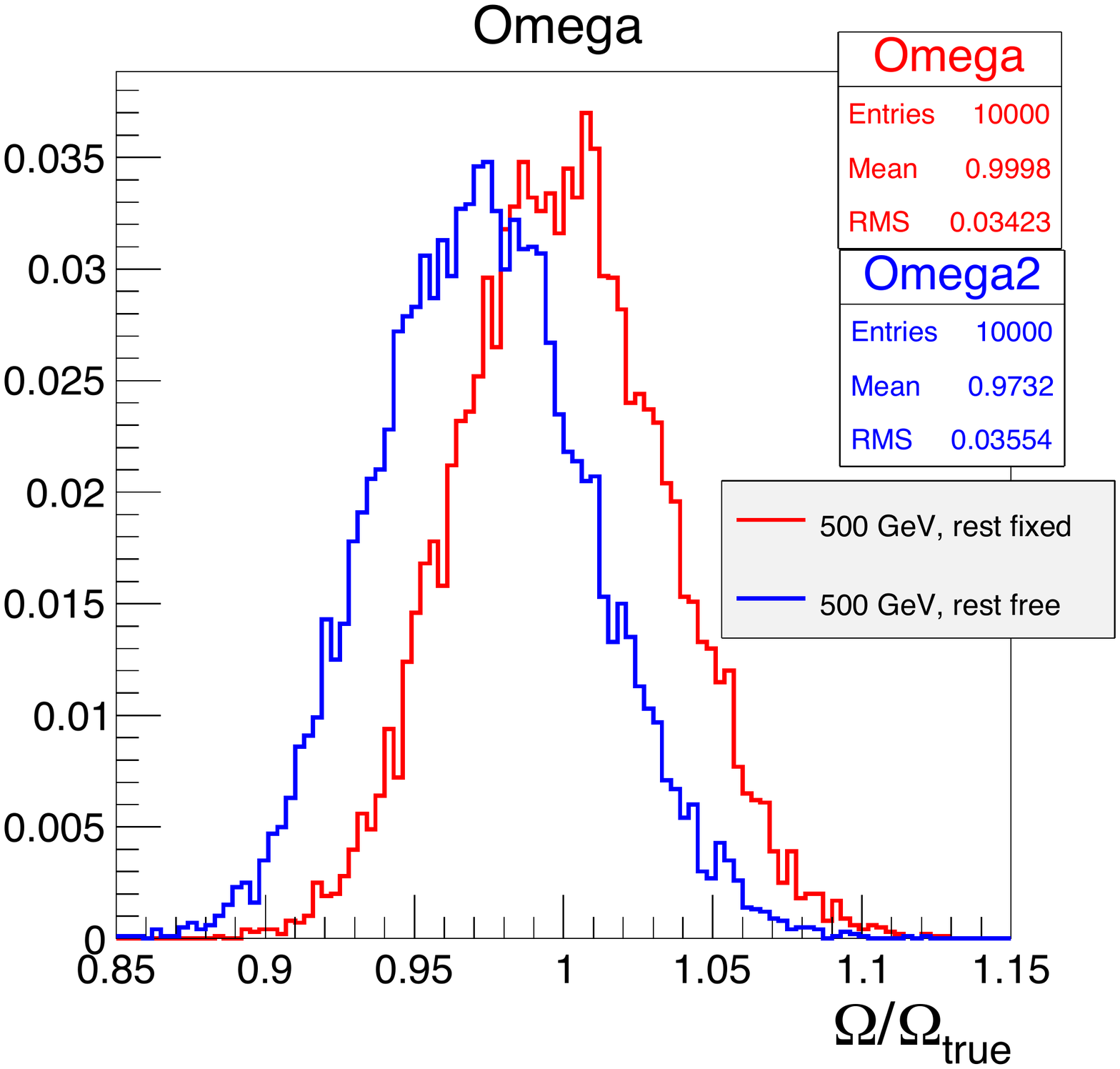}
\caption{The red distribution contains variations of all possible observables from the 500 GeV ILC according to Table \ref{table:ILC500observables} and the rest of the spectrum is fixed. In the blue distribution the same observations are assumed and the rest of the spectrum is allowed to vary uniformly according to Table \ref{table:ILC500notobservables}. This causes a $1\sigma$ shift of the mean predicted value.}
\label{fig:ILC500fixedfree}
\end{center}
\end{figure}

Thus far, true model values have been used for sparticle masses and mixings apart from the LSP and $\stauone$. If the other masses and mixings were allowed to vary, would the relic density still be predicted correctly? This question is answered by Fig. \ref{fig:ILC500fixedfree}. Let us assume that the sleptons, sneutrinos and $\LSP, \neutralinotwo$ and $\charginoone$ were discovered with the ILC at $\sqrt{s}=500$ GeV. Then assuming the uncertainties in Table \ref{table:ILC500observables} and fixing the properties of the squarks, higgsinos and heavy Higgses to the true values, the relic density is predicted correctly with a 3.4\% uncertainty. If the properties of the unobserved squarks, higgsinos and heavy Higgses were allowed to vary uniformly within the ranges in Table $\ref{table:ILC500notobservables}$, then the central value of the dark matter distribution is shifted to the left by about one standard deviation from the true model value. The shift is small, so any possible contribution to the relic density from other types of dark matter would have to be small.

 \begin{table}[ht!]{}
 \centering
$\begin{array}{|  l | l || l |  l  | }
 \hline
 \text{observable} & \text{uncertainty} & \text{observable} & \text{uncertainty} \\ \hline
   \textcolor{blue}{m_{\LSP}} & \textcolor{blue}{0.15\%} & \textcolor{blue}{m_{\neutralinotwo}} & \textcolor{blue}{0.5\%} \\
    \textcolor{blue}{m_{\stauone}} & \textcolor{blue}{0.16\%} & \textcolor{blue}{m_{\stautwo}} & \textcolor{blue}{2.5\%} \\

 \textcolor{blue}{m_{\seR}} & \textcolor{blue}{0.17\%} & \textcolor{blue}{m_{\smuR}} & \textcolor{blue}{0.40\%} \\ 
 m_{\seL} & 1\% & m_{\smuL} & 1\% \\ 

 m_{\snuel, \snumu, \snutau} & 1\% & m_{\charginoone} & 1\%  \\ 

   \theta_{\tau} & 1\% & A_{\tau} & 20\% \\ 
 N_{11, 12, 13, 14} & 1\% \text{ each} & Umix, Vmix & 20\% \text{ each}  \\ \hline
 \end{array}$
\caption{Uncertainties at the ILC with $\sqrt{s}=500$ GeV. Black=estimate, blue=analysis \cite{Berggren:2015qua}.}
\label{table:ILC500observables}
\end{table}

 \begin{table}[ht!]{}
 \centering
$\begin{array}{|  l | l || l  | l  | }
 \hline
  \text{observable} & \text{range} & \text{observable} & \text{range} \\ \hline
  m_{\neutralinothree, \neutralinofour} & 0.25 \to 2 \text{ TeV} & m_{\charginotwo} & 0.25 \to 2 \text{ TeV} \\
 
  m_{H_0, A_0, H^{\pm}} & 0.4 \to 2 \text{ TeV} & &  \\ 

  m_{\widetilde{d}_L, \widetilde{u}_L, \widetilde{s}_L, \widetilde{c}_L} \text{ all equal} & 1 \to 50\text{ TeV} & m_{\widetilde{d}_R, \widetilde{u}_R, \widetilde{s}_R, \widetilde{c}_R} & = m_{\widetilde{d}_L}-100\GeV \\ 
  m_{\widetilde{t}_1, \widetilde{t}_2, \widetilde{b}_1, \widetilde{b}_2} \text{ independent} & 0.6-50\text{ TeV} & m_{\widetilde{g}} & 1-50 \text{ TeV}  \\ 
\theta_{t,b} & -\pi/2 \to \pi/2 & A_{t, b} & -5000 \to 5000 \\  \hline

\end{array}$
\caption{STC8 particles not observed at the ILC with $\sqrt{s}=500$ GeV. All variables are varied uniformly within the indicated ranges.}
\label{table:ILC500notobservables}
\end{table}

\begin{figure}[ht!]
\begin{center}
\leavevmode
\includegraphics[width=0.6\textwidth]{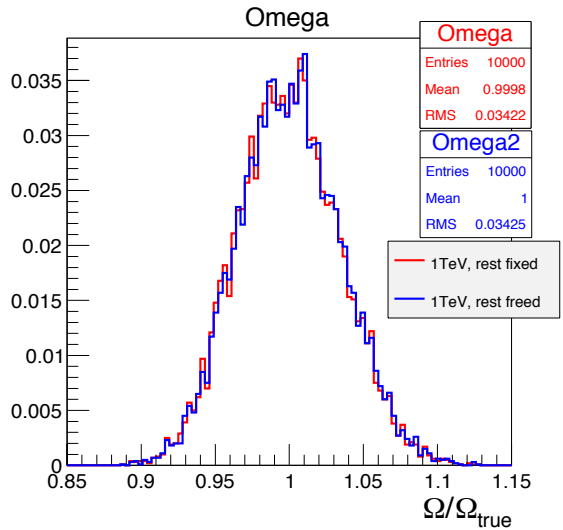}
\caption{The assumptions used for this figure are listed in Tables \ref{table:ILC1000observables} and \ref{table:ILC1000notobservables}. All sparticles except the squarks are assumed to be observed. The relic density is predicted to have the true value with a 3.4\% uncertainty irrespective of whether the squarks are observed (red) or not (blue).}
\label{fig:ILC1000fixedfree}
\end{center}
\end{figure}

The shift would be removed by the observation of the heavy Higgses and the $\neutralinothree, \neutralinofour$ and $\charginotwo$. These would be pair produced and hence observed at the ILC with $\sqrt{s}= 1$ TeV. Some could be observed in mixed production already at $\sqrt{s} = 550$ GeV. Assuming a 1\% uncertainty on the masses of the heavy Higgses, $\neutralinothree, \neutralinofour$ and $\charginotwo$, and keeping the previously used assumptions for the lighter sparticles, the relic density distribution has the same uncertainty as before, as can be seen in Fig. \ref{fig:ILC1000fixedfree}. All of the assumptions are listed in Tables $\ref{table:ILC1000observables}$ and $\ref{table:ILC1000notobservables}$. It does not matter whether the squark masses are varied or not. The centre of the relic density distribution has the true value in both cases and the uncertainty is the same as before. The uncertainty is still dominated by the 1\% precision on $N_{11}$.

It was conservatively assumed that the precisions on the light particles would not improve with the 1 TeV running. This is not the case in reality: the precisions on masses would improve.  In addition, the discovery of the full neutralino sector would decrease the uncertainty on the mixing properties of the LSP, which is the largest contributor to the relic density uncertainty. Currently there does not exist an analysis on the precisions of the 1 TeV measurements.

An observable that was not considered is the mixing angle of the CP-even Higgses. This would be well constrained after the discovery of the heavy Higgses and the determination of $\tan \beta$ from the neutralino or chargino sector. Also the mixings of the $\neutralinotwo$, $\neutralinothree$ and $\neutralinofour$ were fixed to their true values. This was done because the mixing properties of the neutralinos are interlinked by the orthogonality of the neutralino mixing matrix.

 \begin{table}[ht!]{}
 \centering
$\begin{array}{|  l | l || l  | l  | }
 \hline
  \text{observable} & \text{uncertainty} & \text{observable} & \text{uncertainty} \\ \hline
 m_{\seR} & 0.17\% & m_{\smuR} & 0.40\% \\ 
 m_{\seL} & 1\% & m_{\smuL} & 1\% \\ 
 m_{\stauone} & 0.16\% & m_{\stautwo} & 2.5\% \\ 
 \theta_{\tau} & 1\% & A_{\tau} & 20\% \\ 
 m_{\snuel, \snumu, \snutau} & 1\% & m_{\charginoone} & 1\%  \\ 
  m_{\LSP} & 0.15\% & m_{\neutralinotwo} & 0.5\% \\ 
 N_{12, 13, 14} & 1\% \text{ each} & Umix, Vmix & 20\% \text{ each}  \\ 
  m_{\neutralinothree, \neutralinofour} & 1\% & m_{\charginotwo} & 1\% \\
 
  m_{H_0, A_0, H^{\pm}} & 1\% & &  \\ \hline
 \end{array}$
\caption{STC8 particles observed at the 1 TeV ILC}
\label{table:ILC1000observables}
\end{table}

  \begin{table}[ht!]{}
  \centering
$\begin{array}{|  l | l || l  |   l   |}
\hline
  \text{observable} & \text{range} & \text{observable} & \text{range} \\ \hline
  m_{\widetilde{d}_L, \widetilde{u}_L, \widetilde{s}_L, \widetilde{c}_L} \text{ all equal} & 1\to50\text{ TeV} & m_{\widetilde{d}_R, \widetilde{u}_R, \widetilde{s}_R, \widetilde{c}_R} & = m_{\widetilde{d}_L}-100\GeV \\ 
  m_{\widetilde{t}_1, \widetilde{t}_2, \widetilde{b}_1, \widetilde{b}_2} \text{ independent} & 0.6\to50\text{ TeV} & m_{\widetilde{g}} & 1\to50 \text{ TeV}  \\ 
\theta_{t,b} & 0 \to \pi/2 & A_{t, b} & 0 \to -5000 \\ \hline

\end{array}$
\caption{STC8 particles not observed at the 1 TeV ILC}
\label{table:ILC1000notobservables}
\end{table}

A limitation of this study is that \pkg{micrOMEGAs} calculates the SUSY cross sections at tree-level. This means that the importance of any unobserved particles is underestimated. Loop corrections to stau coannihilation diagrams can change the cross section by $\sim10\%$ \cite{Baro:2007em}. The full set of one-loop SUSY corrections has been calculated with the program \pkg{SLOOPS} \cite{Baro:2007em}. It would be possible to interface \pkg{SLOOPS} with \pkg{micrOMEGAs} to improve improve the accuracy of the study.

\section{Conclusions}

The ILC could discover SUSY scenarios with small mass differences. Scenarios with a stable bino LSP and an almost mass degenerate $\stau$, $\widetilde{t}$ or $\chargino$ are among possible candidates which could explain the dark matter relic density that the Planck mission has measured. This study investigated a stau coannihilation benchmark point where all sleptons, sneutrinos, $\LSP$, $\neutralinotwo$ and $\charginoone$ could be discovered at the ILC at $\sqrt{s}=500$ GeV. Assuming these discoveries, the mixings of the $\stauone$ and the LSP would have to be measured to the percent-level precision to predict the relic density with an accuracy comparable to the current cosmological accuracy. The main mixing of the LSP dominates the uncertainty on the relic density prediction.  The squarks and the gluino do not have to be observed.  Measuring the full neutralino and Higgs sectors with the ILC at $\sqrt{s}=1$ TeV would confirm or exclude the existence of other types of dark matter than the LSP.

\bibliography{darkmatter2}

\end{document}